\documentclass[fleqn,usenatbib]{mnras}

\usepackage{graphicx}	
\usepackage{amsmath}	
\usepackage{amssymb}	
\usepackage{multicol}        

\usepackage{color}
\usepackage{tcolorbox}
\usepackage{pdflscape}
\usepackage[T1]{fontenc}
\usepackage{ae,aecompl}
\usepackage{newtxtext,newtxmath}
\usepackage{lscape}
\usepackage{mathtext}
\usepackage{float}
\newcommand{\beq}{\begin{equation}}
\newcommand{\eeq}{\end{equation}}
\newcommand{\beqa}{\begin{eqnarray}}
\newcommand{\eeqa}{\end{eqnarray}}

\newcommand{\SRG}{{\it SRG}}
\newcommand{\ART}{{\it ART-XC}}
\newcommand{\eROSITA}{{\it eROSITA}}
\newcommand{\Chandra}{{\it Chandra}}
\newcommand{\XMM}{{\it XMM-Newton}}
\newcommand{\WISE}{{\it WISE}}

\newcommand{\cfhqs}{CFHQS\,J142952+544717}
\newcommand{\srge}{SRGE\,J142952.1+544716}
\newcommand{\pso}{PSO\,J030947.49+271757.31}

\newcommand{\ar}{\alpha_{\rm r}}
\newcommand{\aosx}{\alpha_{\rm ox}}
\newcommand{\aox}{\tilde{\alpha}_{\rm ox}}

\newcommand{\nuo}{\nu_{2500\,\textup{\AA}}}
\newcommand{\Lo}{L_{2500\,\textup{\AA}}}
\newcommand{\nux}{\nu_{10\,{\rm keV}}}
\newcommand{\Lx}{L_{10\,{\rm keV}}}
\newcommand{\nusx}{\nu_{2\,{\rm keV}}}
\newcommand{\Lsx}{L_{2\,{\rm keV}}}
\newcommand{\Lbol}{L_{\rm bol}}
\newcommand{\Luv}{L_{<2\,{\rm keV}}}
\newcommand{\Lbx}{L_{2-30\,{\rm keV}}}
\newcommand{\Lex}{L_{2-100\,{\rm keV}}}

\title{SRG/eROSITA uncovers the most X-ray luminous quasar at $z>6$}  
\author[Medvedev et al.] {Medvedev P.$^1$,
Sazonov S.$^{1,2}$,
Gilfanov M.$^{1,3}$,
Burenin R.$^1$,
Khorunzhev G.$^1$,
\newauthor
Meshcheryakov A.$^{1,4}$,
Sunyaev R.$^{1,3}$,
Bikmaev I.$^{4,5}$,
Irtuganov E.$^{4,5}$\\
$^1$Space Research Institute (IKI), Russian Academy of Sciences, Profsoyuznaya ul. 84/32, Moscow, 117997 Russia \\
$^2$Moscow Institute of Physics and Technology, Institutsky per. 9, 141700 Dolgoprudny, Russia \\
$^3$Max-Planck-Institut f\"{u}r Astrophysik (MPA), Karl-Schwarzschild-Str. 1, D-85741 Garching, Germany \\
$^4$Kazan Federal University, Kremlevskaya str.18, 420008 Kazan, Russia\\
$^5$Academy of Sciences of Tatarstan, Baumana Str., 20, 420111, Kazan, Russia
}

\begin{document}

\maketitle

\begin{abstract}
 We report the discovery of X-ray emission from \cfhqs, the most distant known radio-loud quasar at $z=6.18$, on Dec. 10--11, 2019 with the \eROSITA\ telescope on board the \SRG\ satellite during its ongoing all-sky survey. The object was identified by cross-matching an intermediate \SRG/\eROSITA\ source catalog with the Pan-STARRS1 distant quasar sample at $5.6 < z < 6.7$. The measured flux $\sim 8 \times 10^{-14}$~erg~cm$^{-2}$~s$^{-1}$ in the 0.3--2~keV  energy band corresponds to an X-ray luminosity of $2.6^{+1.7}_{-1.0}\times 10^{46}$~erg~s$^{-1}$ in the 2--10~keV rest-frame energy band, which renders \cfhqs\ the most X-ray luminous quasar ever observed at z $> 6$. Combining our X-ray measurements with archival and new photometric measurements in other wavebands (radio to optical), we estimate the bolometric luminosity of this quasar at $\sim (2$--$3) \times 10^{47}$~erg~s$^{-1}$. Assuming Eddington limited accretion and isotropic emission, we infer a lower limit on the mass of the supermassive black hole of $\sim 2\times 10^9~M_\odot$. The most salient feature of \cfhqs\ is its X-ray brightness relative to the optical/UV emission. We argue that it may be linked to its radio-loudness (although the object is not a blazar according to its radio properties), specifically to a contribution of inverse Compton scattering of cosmic microwave background photons off relativistic electrons in the jets. If so, \cfhqs\ might be the tip of the iceberg of high-$z$ quasars with enhanced X-ray emission, and \SRG/\eROSITA\ may find many more such objects during its 4~year all-sky survey. 
\end{abstract}


\section{Introduction}
\label{s:intro}

One of the most intriguing questions in astrophysics today is how the first supermassive black holes (SMBHs) appeared in the early Universe and how their growth was interconnected with the formation and evolution of the first galaxies. Recently, the number of known quasars at $z>6$ (i.e. when the Universe was younger than 950 million years) has increased dramatically to $\sim 200$ thanks to the advent of wide-area optical and IR surveys (in particular, the CFHT Legacy Survey, CFHTLS, \citealt{Willott_CFHQS}; Pan-STARRS, \citealt{ps1_survey}; the Dark Energy Survey, DES, \citealt{Yang_distant_quasars}; see also \citealt{Banados_distant_quasars}). By analogy with the much better studied active galactic nuclei (AGNs) at lower redshift, one can expect a substantial fraction of the bolometric luminosity of distant quasars to be emitted in the X-ray band. However, so far only $\approx 35$, i.e. $\sim 15$\%, of the known $z>6$ quasars have been observed in X-rays (in dedicated or serendipitous observations, mostly by \Chandra\ or \XMM). As a result, $\approx 21$ quasars have been detected in X-rays and upper limits on the X-ray luminosity have been obtained for another $\approx 14$ objects (see \citealt{Vito_xray_quasars, pons_quasars} and references therein).  

All of the aforementioned $z>6$ quasars observed in X-rays so far are radio-quiet (RQQs), which largely reflects the paucity of known radio-loud quasars (RLQs) at high redshift. Specifically, until recently only 9 (3) RLQs were known at $z>5.5$ ($z>6$) \citep{Banados_radioloud_quasars,Belladitta_blazar}\footnote{Hereafter, we use the traditional dividing line $R=10$ between RQQs and RLQs, where $R\equiv f_{\nu,\,5\,GHz}/f_{\nu,\,4400\AA}$, with $f_{\nu,\,5\,GHz}$ and $f_{\nu,\,4400\AA}$ being the flux densities at rest-frame 5~GHz and 4400~\AA.}, and none of the three $z>6$ RLQs have been observed in X-rays so far. Recently, \cite{Belladitta_blazar} have discovered a 4th RLQ at $z>6$, \pso, with $z=6.10\pm0.03$, by cross-matching the catalogs of sources detected in the NVSS (radio), Pan-STARRS (optical) and WISE (mid-infrared) surveys. This object proved to be extremely interesting, as it is not just radio-loud but a blazar, and in fact the most distant blazar known. Following the discovery, \cite{Belladitta_blazar} carried out X-ray observations of \pso\ with {\it Neil Gehrels Swift/XRT} and detected a signal corresponding to a luminosity of $4.4^{+6.4}_{-3.0}\times 10^{45}$~erg~s$^{-1}$ in the rest-frame 2--10~keV energy band, possibly (taking into account the large uncertainty in the luminosity) making this object the most X-ray luminous quasar known at $z>6$. 

On July 13, 2019, the \SRG\ X-ray observatory (Sunyaev et al., 2020, in preparation) was launched from the Baikonur cosmodrome. It carries two wide-angle grazing-incidence X-ray telescopes, \eROSITA\ (Predehl et al., 2020, in preparation) and \ART\ (Pavlinsky et al., 2020, in preparation), operating in the overlapping energy bands of 0.3--10 and 4--30~keV, respectively. On Dec. 8th, 2019, upon completion of the commissioning, calibration and performance verification phases, \SRG\ started its all-sky X-ray survey from a halo orbit around the Sun--Earth L2 point, which will consist of 8 repeated 6~month long scans of the entire sky. Already a preliminary analysis of the data accumulated during the first months of this survey has demonstrated its enormous discovery potential. 

Here we report the first detection of X-ray emission from \cfhqs, the most distant ($z=6.18$) known RLQ, with \SRG/\eROSITA. This object turns out to be the most X-ray luminous quasar known at $z>6$ and in fact even more luminous than the blazar \pso\ discussed above. The fact that \cfhqs\ does not appear to be a blazar according to its properties in the radio makes this discovery even more interesting. In what follows we adopt a flat $\Lambda$CDM cosmological model with $h = 0.70$ and $\Omega_{\Lambda} = 0.7$. All quoted uncertainties correspond to 1$\sigma$, unless noted otherwise.

\section{X-ray observations}
\label{s:xray}

\begin{figure*}
\includegraphics[width=1\textwidth]{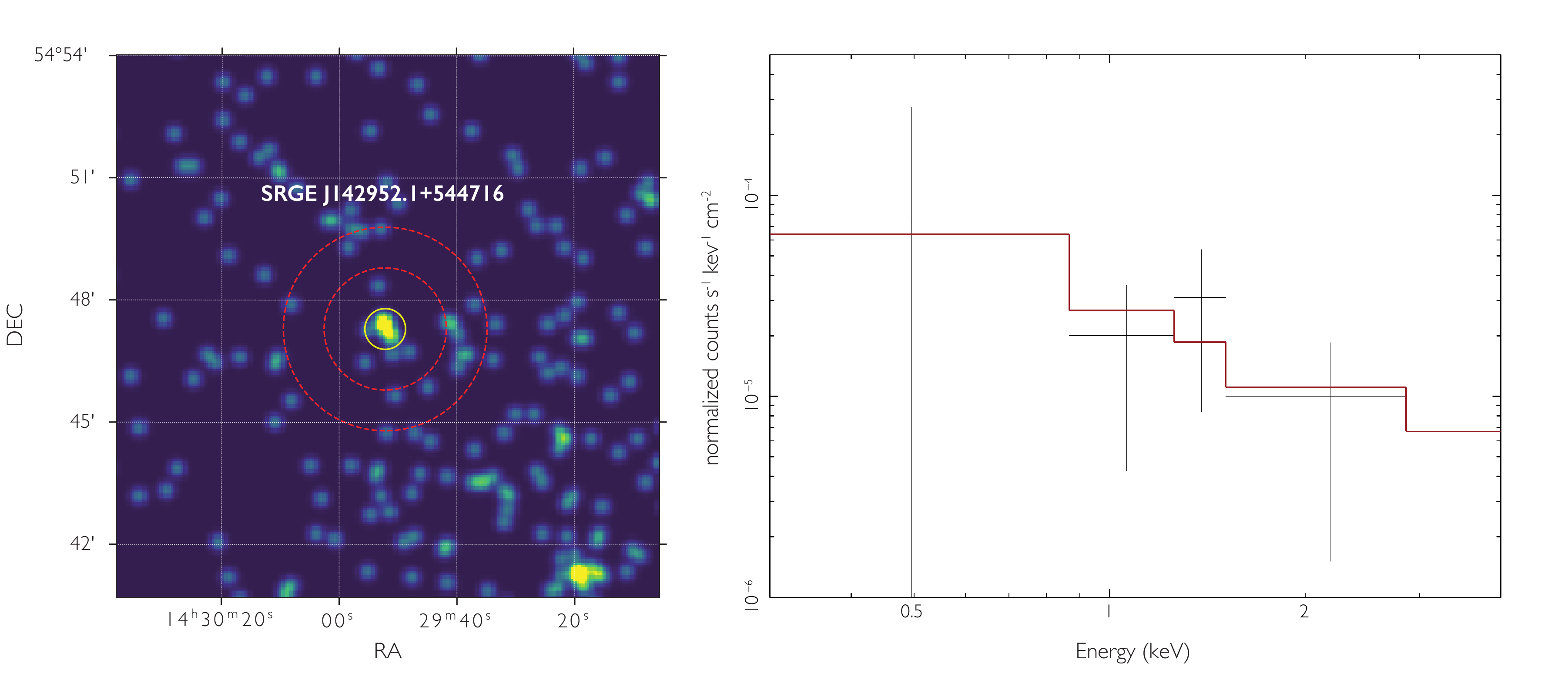}
\caption{Quasar \cfhqs\ as seen by \SRG/\eROSITA\ on Dec. 10--11, 2019. Left: $13^{\prime}\times13^{\prime}$ X-ray (0.3--2~keV) image centered at \cfhqs\ optical coordinates,  smoothed with a Gaussian filter with $\sigma=8^{\prime\prime}$. The yellow solid circle depicts the source extraction region with a radius of 30~arcsec and the red dashed circles show the annulus used for background extraction (90 to 150~arcsec). Right: X-ray spectrum from the source extraction region (black) and the absorbed power-law model providing the best fit to the data in the 0.3--4~keV energy band (red). The energy channels are rebinned to ensure at least 1$\sigma$ significance in each bin.} 
\label{fig:xray}
\end{figure*}

\cfhqs\ was found by cross-matching an intermediate release of the \eROSITA\ all-sky survey source catalog of the Eastern Galactic hemisphere (where data analysis lies within the responsibility of the Russian consortium) with the Pan-STARRS1 (PS1) distant quasar sample of 124 quasars in the redshift range of $5.6 < z < 6.7$ \citep{Banados_distant_quasars}. 

The position of \cfhqs\ was scanned by the \SRG\ observatory 8 times during Dec. 10--11, 2019. On Dec. 10, the scans were interrupted by an orbital correction of the spacecraft. During the correction manoeuvre and for several hours after it, \eROSITA\ was not collecting X-ray photons.  The total vignetting corrected exposure time on the source is thus about 160~s.

The \eROSITA{} raw data were processed by the calibration pipeline based on the eROSITA Science Analysis Software System (eSASS) and using pre-flight calibration data. A new X-ray source \srge{} was detected at the position 
$\alpha=14^{h}29^{m}52.14^{s}$, $\delta=+54^{d}47^{m}15.7^{s}$
with a 1$\sigma$ positional uncertainty of 4.5~arcsec as calculated by the {\it ermldet} task of eSASS. The detection likelihood is 20.1 in the 0.3--2.0~keV band, i.e. the detection is highly significant. The best-fit X-ray source position deviates by 1.9~arcsec from the optical position of the quasar \cfhqs{} ($\alpha=14^{h}29^{m}52.146^{s}$, $\delta=+54^{d}47^{m}17.55^{s}$) 
in CFHTLS \citep{Hudelot_CFHTLS}.  Given the average \eROSITA\ source density in this field of  $\sim 5$ sources per deg$^2$, the probability of finding by chance an X-ray source within 1.9~arcsec from the quasar's optical position is $\sim 4.4\times 10^{-6}$. Taking into account the number of optical quasars from \citet{Banados_distant_quasars} within the footprint of the \eROSITA\ all-sky survey at the time of this analysis, which were checked against the X-ray catalog (45 sources), the total probability of chance coincidence of an \eROSITA\ source and at least one of these $z\sim 6$ quasars is $\sim 2\times 10^{-4}$.

For further analysis, we chose an extraction region centered at the quasar's optical position with a radius of 30~arcsec, which is approximately twice the half-energy radius of the \eROSITA\ survey's point spread function in the 0.3--2.0~keV energy band \citep{erosita_science}. In this region, we detected a total of 9~counts, with the expected background contribution of $\approx 0.8$~counts. The X-ray image with the superimposed source and background regions is shown in Fig.~\ref{fig:xray} (left panel).  The probability to register $\ge 9$ counts given the Poisson distribution with a mean of $\lambda=0.8$ is $p\approx 1.4\times  10^{-8}$; $-\ln(p)=18.1$ is roughly consistent with the detection likelihood of $20.1$ quoted above. This confirms the high statistical significance of the X-ray source detection. Multiplying this number by the number of optical quasars from \citet{Banados_distant_quasars} within the footprint of the \eROSITA\ all-sky survey, we find that the probability of a fluctuation of this magnitude is $\sim 6\times 10^{-7}$. We thus conclude that the detection of X-ray emission from \cfhqs\ is highly reliable.

Because of the small number of detected photons, only crude spectral analysis is feasible. To roughly characterise the source's spectrum, we fit it with an absorbed power law with the neutral hydrogen column density fixed at the Galactic value in the direction of \cfhqs\, $N_{\rm H}=1.15\times10^{20}$~cm$^{-2}$, obtained from the HI4PI map \citep{HI4PI_collab}. 
 For the spectral fit we consider the 0.3--4~keV energy band since no counts have been detected
within the extraction region at higher energies. Using the C-statistic \citep{cstat}, we obtain the best-fit photon index $\Gamma=1.4\pm 0.9$ and the absorption corrected fluxes $F_{0.3-2} = 8.2^{+3.7}_{-2.7} \times 10^{-14}$~erg~cm$^{-2}$~s$^{-1}$  and $F_{2-4} = 6.2^{+8.9}_{-3.9} \times 10^{-14}$~erg~cm$^{-2}$~s$^{-1}$ in the 0.3--2 and 2--4~keV energy bands (observer's frame), respectively (see Fig.~\ref{fig:xray}). Fixing the photon index at $\Gamma=1.8$ (as is typical of AGNs), we obtain  similar  values: $F_{0.3-2} = 8.8^{+3.5}_{-2.8} \times 10^{-14}$~erg~cm$^{-2}$~s$^{-1}$ and $F_{2-4} = 4.1^{+1.7}_{-1.3} \times 10^{-14}$~erg~cm$^{-2}$~s$^{-1}$. No statistically significant variability was found over the period of observations (Dec. 10--11, 2019), although the total number of photons detected from the source is too low for a meaningful variability analysis. 

Adopting the quasar's redshift $z=6.183$ from the CO (2--1) line observations of its host galaxy (\citealt{Wang_CO}, which is slightly different from the estimate $z=6.21$ based on the position of the Ly$\alpha$ emission line in the quasar's spectrum, \citealt{Willott_CFHQS}) and using our best-fit spectral model, we can estimate the X-ray luminosity of \cfhqs\ at $2.6^{+1.7}_{-1.0} \times 10^{46}$~erg~s$^{-1}$ in the rest-frame 2--10~keV energy band.

 
\section{Multiwavelength properties}

Apart from its detection in X-rays by \SRG/\eROSITA, \cfhqs\ is well studied in the radio, mid-infrared and optical wavebands. Below we discuss new and previously published multi-wavelength data  available for \cfhqs\ (see Table~\ref{tab:sed} for a compilation of all measurements).

\subsection{Optical}

In the optical, with absolute magnitude (at rest-frame 1450\,\AA) $M_{1450}\approx{}-26.0\pm0.1$ (\citealt{Willott_CFHQS,Omont_fir,Banados_distant_quasars}), \cfhqs\ belongs to the bright end of the quasar luminosity function at $z\sim 6$ (e.g. \citealt{Willott_CFHQS}), but is not unique, being $\sim 1.5$ magnitudes fainter than the most optically luminous quasars known at these redshifts. 

The object was discovered on 6 Dec 2010 during the Canada--France High-$z$ Quasar Survey (CFHQS), where its magnitudes $i^{\prime} = 23.88\pm0.06$ and $z^{\prime}= 21.45 \pm 0.03$ were measured \citep{Willott_CFHQS}. The $u^{\prime}$$g^{\prime}$$r^{\prime}$$i^{\prime}$$z^{\prime}$ photometry was updated in the CFHTLS-Wide survey catalog \citep{Hudelot_CFHTLS}. Later, the source was also detected in the CFHQSIR survey (a Y-band extension of CFHTLS-Wide survey, 2010--2012, \citealt{CFHQSIR}), with $Y_{\rm AB}=20.64\pm0.1$ (Kron-like aperture), and in the PS1 survey \citep{ps1_survey_16}, with $z_{\rm PS1, PSF}=21.94\pm 0.13$. 

\subsubsection{Serendipitous source CFHT\,1323\_1754990}

\begin{figure*}
\centering
\includegraphics[width=0.9\textwidth]{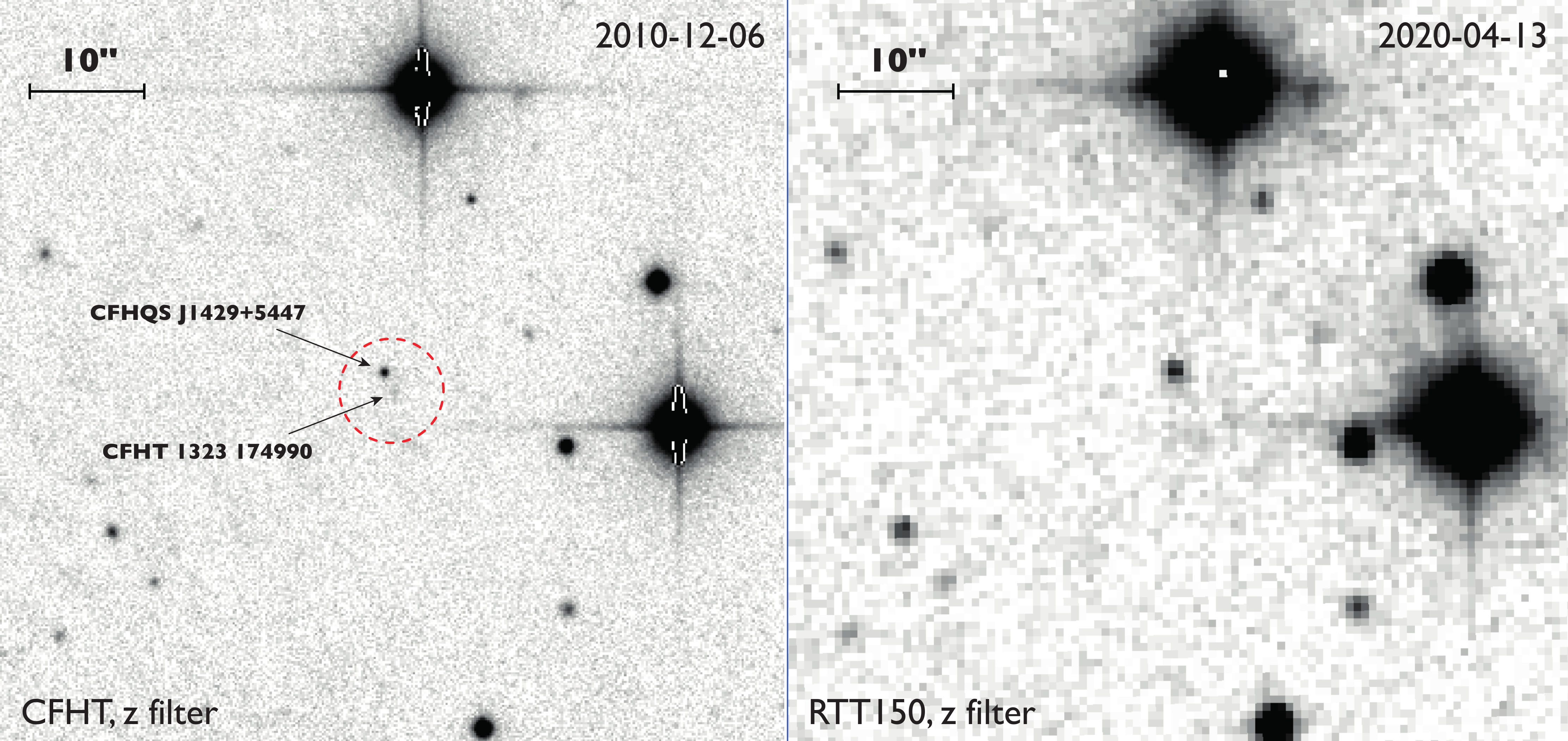}
\caption{Left: archival image of the \cfhqs\ field obtained from the CFHTLS-Wide public data archive (http://www.cadc-ccda.hia-iha.nrc-cnrc.gc.ca) in the $z$ band. The red dashed circle shows the 1$\sigma$ positional uncertainty (4.5 arcsec) of \srge\ and is centered at its position. Also marked is the position of another faint optical source, CFHT\,1323\_1754990, at 2~arcsec from \cfhqs. Right: image of the same field in the $z$ band obtained with RTT150 on April 13, 2020.
}

\label{fig:optic}
\end{figure*}

In the CFHTLS-Wide survey catalog \citep{Hudelot_CFHTLS}, there is another faint optical object (hereafter source \#2) within the 68\% error circle of the X-ray source, at just 1.7~arcsec from the quasar \cfhqs\ (see Fig.~\ref{fig:optic}). The object, designated as CFHT\,1323\_1754990 in the CFHTLS catalog, is located at  $\alpha=14^{h}29^{m}52.085^{s}$, $\delta=+54^{d}47^{m}15.92^{s}$,  and has the following magnitudes (in a Kron-like aperture):
$u^{\prime}=24.02\pm0.09$,
$g^{\prime}=24.15\pm0.06$,
$r^{\prime}=24.38\pm0.14$,
$i^{\prime}=23.81\pm0.13$ and 
$z^{\prime}=22.23\pm0.09$.
Hence, source \#2 is similar in brightness to \cfhqs\ in the $i^{\prime}$ band but has a much bluer optical spectral energy distribution, typical for field galaxies. 

The probability of finding such an optical object in a circle of radius of $4.5\arcsec$ (the 68\% error circle of the X-ray source) by chance is $p_1\sim 0.2$. It is high because the number density of such faint objects (at least as bright as source \#2) is quite high, $\sim 4\times 10^4$ sources~deg$^{-2}$, as determined from the CFHTLS catalog.  

Conversely, it is also possible that the X-ray emission is actually associated with source \#2, while the quasar is projected by chance. However, this is quite unlikely because bright X-ray sources like \srge\ are rarely associated with faint optical objects like source \#2. To demonstrate this, we inspected the W1 field (65~deg$^2$) of CFHTLS-Wide located inside the footprint of the intermediate \eROSITA\ source catalog and having exposure time similar to that for \srge. We obtained a conservative upper limit of a few per cent on the occurrence of X-ray bright, $F_{\rm X}> 8\times 10^{-14}$~erg~cm$^{-2}$~s$^{-1}$, sources with optical counterparts as bright or fainter than source \#2, yielding an upper limit on their surface density near \cfhqs\ of 0.2~sources~deg$^{-2}$. Therefore, the probability that one of the 45 distant quasars from \citet{Banados_distant_quasars} is found by chance within 1.7~arcsec from an object like source \#2, is  $p_2< 7\times 10^{-6}$.

As $p_2\ll p_1$ and $p_2< 7\times 10^{-6}$, it is highly unlikely that \srge\ is associated with CFHT\,1323\_1754990 rather than with the quasar \cfhqs. We thus exclude this possibility from further consideration.

We finally note that visual inspection of the CFHTLS images reveals another possible faint optical source within less than one arcsec from \cfhqs. It is not listed in existing optical catalogs and at least one magnitude fainter than source \#2. Because of its faintness, the probability that the X-ray emission detected by \eROSITA\ is associated with this source is yet lower that for CFHT\,1323\_1754990.



%

\subsubsection{RTT150 observations}

Following the detection of X-ray emission from \cfhqs\ by \SRG/\eROSITA, we carried out photometric observations at the Russian--Turkish 1.5-m Telescope (RTT150) of the TUBITAK National Observatory, Antalya, Turkey, in order to check a possible brightening of the quasar in the optical. The observations took place on two nights: on March 8, 2020 (i.e. almost 3 months after the X-ray detection) and on April 13, 2020.

The images were obtained with the TUBITAK Faint Object Spectrograph and Camera (TFOSC) instrument equipped with an ANDOR DZ936 BR DD CCD. For the first observation run we took a series of 6$\times$600\,s images in the SDSS-$i$ filter and 4$\times$600\,s images in the SDSS-$z$ filter. For the second run, a series of $8\times$600\,s images in the SDSS-$z$ filter were obtained. The combined images were photometrically calibrated using the magnitudes of stars in the observed field taken from the SDSS DR14 database \citep{sdss_dr14}. 

We measured the source's brightness at $z^{\prime}=21.20\pm0.12$ on March 8 and at $z^{\prime}=21.31\pm0.06$ on April 13 (see Fig.~\ref{fig:optic}). We detected no flux in the $i$ filter with an upper limit of $i^{\prime} > 23.5$ ($5\sigma$). We estimate the magnitude corrections associated with the differences in the $z$ band between the RTT150 observation and the surveys mentioned above to be smaller than $\Delta z^{\prime} \approx 0.1$. Therefore, the new measurements of the $z^{\prime}$ magnitude are in approximate agreement with previous measurements, whereas our lower limit on the $i^{\prime}$ magnitude is consistent with the corresponding measurement in CFHTLS-Wide.

We conclude that photometric measurements taken on a number of occasions over the past $\sim 10$ years do not reveal a substantial (by more than a factor of $\sim 2$) variability of \cfhqs\ in the optical band, with the caveat that \SRG/\eROSITA\ may have caught the source in an outburst in Dec. 2019 with no contemporaneous optical observations being available for that period. 


\subsection{Mid-infrared}
\cfhqs\ has been detected by \WISE\ \citep{wright10,cutri13} in the mid-infrared, however only the $W1$-band (3.4~$\mu$m) measurements are available in the AllWISE Data Release with sufficient detection significance. We took advantage of a ``forced photometry'' technique to estimate the $W1$ and $W2$ magnitudes  using the NEOWISE 2019 Data Release \citep{mainzer14} and the  precise position of \cfhqs\ provided by PS1. We thus obtained the source's magnitudes $W1=17.24$ and $W2=16.69$ (in the Vega system).

\subsection{Radio}
\label{s:radio}

In the radio, \cfhqs\ has been detected at several frequencies between 1.4 and 32~GHz in wide-angle surveys (FIRST and NVSS, \citealt{Helfand_first,Condon_nvss}) and dedicated observational campaigns \citep{Wang_CO,Frey_VLBI}. In particular, \cite{Frey_VLBI} performed VLBI imaging at 1.6 and 5~GHz and inferred that the radio source (i) is compact but resolved on linear scales $<100$~pc, (ii) has a steep spectrum with $\ar\approx 1.0$\footnote{Hereafter, we define $\ar$ so that $F_\nu\propto \nu^{-\ar}$.} and (iii) relatively low rest-frame brightness temperature $T_{\rm b}\sim 10^9$~K. Having $\ar> 0.5$, \cfhqs\ is confidently classified as a steep-spectrum source \citep{Coppejans_radio}. Furthermore, \cfhqs\ does not appear to be strongly variable in the radio, as suggested by similar flux densities (within a factor of 1.5) at $\sim 1.5$~GHz measured in different observations separated by years (see fig.~14 in \citealt{Coppejans_radio} and Table~\ref{tab:sed}). All these properties imply that relativistic beaming does not play a significant role in the appearance of \cfhqs, hence {\it it is not a blazar}. 

Recently, \cfhqs\ was detected in the LOFAR Two-meter Sky Survey in the low-frequency band of 120--168~MHz, with a flux density $F_\nu\approx 10$~mJy \citep{Shimwell_lofar}. Comparing this measurement with the aforementioned measurements at higher frequencies, we infer that the spectrum flattens below $\sim 1$~GHz from $\ar\sim 1$ to $\ar\sim 0.5$ (see Table~\ref{tab:sed}). This most likely indicates that the spectrum eventually turns over at $\nu\lesssim 100$~MHz (i.e. at $\nu\lesssim 1$~GHz in the quasar's rest frame), as is expected to occur due to synchrotron self-absorption (see e.g. \citealt{Coppejans_radio}).

\subsection{Microwave (the host galaxy)}

\cfhqs\ has also been investigated at 250~Ghz \citep{Omont_fir}, which corresponds to the far-infrared (FIR) band ($\sim 170$~$\mu$m) in the object's rest frame, and in the CO (2--1) line \citep{Wang_CO}. These observations of the dust and molecular gas revealed that massive star formation (a few hundred $M_\odot$ per year) takes place in this object. In addition, it exhibits a two-component structure in the CO (2--1) image. The position of one component coincides with the optical position of the quasar, while the other component is at a distance of 1.2\arcsec\ (6.9~kpc), with both being at the same redshift ($z\approx 6.1834$) within the uncertainties. It is thus possible that we are witnessing the merger of two galaxies. As discussed by \cite{Wang_CO}, similar pairs have also been observed in other distant quasars. 

All these observational facts clearly draw a picture of violent star formation and rapid SMBH growth taking place hand in hand in \cfhqs. 

\section{Spectral energy distribution}
\label{s:sed}

\begin{table*}
  \caption{Multifrequency properties of \cfhqs. 
    \label{tab:sed}
  }
  \begin{tabular}{|l||l|l|r|r|r|}
    \hline
    \multicolumn{1}{|c|}{Telescope/Survey} &    
    \multicolumn{1}{|c|}{Band} &
    \multicolumn{1}{c|}{Flux or magnitude} &
    \multicolumn{1}{c|}{Ref.} & 
    \multicolumn{1}{c|}{Rest-frame $\nu$ (Hz)} & 
    \multicolumn{1}{c|}{$\nu L_{\nu}$, erg~s$^{-1}$} \\
    \hline
    \multicolumn{6}{c}{Radio:}\\
    LOFAR & 120--168~MHz & $9.8\pm 0.2$ mJy & 1 & $1.0\times 10^{9}$ & $(6.0\pm0.1) \times 10^{42}$\\
    NVSS & 1.4~GHz & $3.8\pm 0.5$ mJy & 2 & $1.0\times 10^{10}$ & $(2.3\pm0.3) \times 10^{43}$ \\
    FIRST & 1.4~GHz & $2.95\pm 0.15$ mJy & 3 & $1.0\times 10^{10}$ & $(1.8\pm 0.1) \times 10^{43}$ \\
    VLBI & 1.6~GHz & $3.30\pm 0.06$ mJy & 4 & $1.2\times 10^{10}$ & $(2.26 \pm 0.04) \times 10^{43}$ \\
    & 5~GHz & $0.99\pm 0.06$ mJy & & $3.6\times 10^{10}$ & $(2.1 \pm 0.1) \times 10^{43}$ \\
    EVLA & 32~GHz & $0.257\pm 0.015$ mJy & 5 & $2.3\times 10^{11}$ & $(3.5 \pm 0.2)\times 10^{43}$ \\
    IRAM/MAMBO & 250~GHz & $3.46\pm 0.52$ mJy & 6 & $1.8\times 10^{12}$ & $(3.7 \pm 0.6)\times 10^{45}$\\ 
    \\
    \multicolumn{6}{c}{Mid-infrared:}\\
    WISE (ALLWISE) & $W4_{\rm AB}$ & $>16.1$ & 7 & $9.7\times10^{13}$ & $<8 \times 10^{46}$\\
    & $W3_{\rm AB}$ & $>17.8$ & & $1.9\times10^{14}$ & $<3 \times 10^{46}$ \\
    WISE/PS1 & $W2_{\rm AB, forced}$  & $20.03\pm0.07$ & 8 & $4.7\times10^{14}$ & $(9.9\pm 0.6)\times 10^{45}$ \\
    & $W1_{\rm AB, forced}$ & $19.94\pm0.03$ & & $6.3\times10^{14}$ & $(1.46\pm 0.05)\times 10^{46}$ \\
    \\
    \multicolumn{6}{c}{Optical:}\\
    PS1 & $y$ & $20.90\pm0.15$ & 9 & $2.2\times10^{15}$ & $(2.15\pm 0.3)\times 10^{46}$ \\
    & $z$ & $21.94\pm0.13$ & & $2.5\times10^{15}$ & $(9.24\pm 1.1)\times 10^{45}$ \\
    & $i$ & $23.54\pm0.28$ & & $2.9\times10^{15}$ & $(2.4\pm 0.6)\times 10^{45}$ \\
    & $r$ & $>22.84$ & & $3.5\times10^{15}$ & $<5.7\times 10^{45}$\\
    & $g$ & $>23.42$ & & $4.4\times10^{15}$ & $<4.3\times 10^{45}$\\
    CFHQSIR & $Y$, Kron-like aperture & $20.64\pm0.1$ & 10& $2.2\times10^{15}$ &$(2.63\pm 0.24)\times 10^{46}$\\
    CFHTLS & $z^\prime$, Kron-like aperture & $21.45\pm0.05$ & 11 & $2.3\times10^{15}$ & $(1.35\pm0.06)\times10^{46}$\\
    & $i^\prime$, Kron-like aperture & $23.92\pm0.16$ & & $2.8\times10^{15}$ & $(1.7\pm0.2)\times10^{45}$\\
    & $r^\prime$, Kron-like aperture & $25.82\pm0.56$ & & $3.4\times10^{15}$ & $(3.6\pm1.8)\times10^{44}$\\
    & $g^\prime$, Kron-like aperture & $25.54\pm0.24$ & & $4.5\times10^{15}$ & $(6.3\pm1.4)\times10^{44}$\\
    & $u^\prime$, 3$\arcsec$-aperture & $>26.6$ & & $6.1\times10^{15}$ & $<3.2\times10^{44}$ \\
    RTT150 & $z^\prime$ & $21.31\pm0.06$ & 8 & $2.4\times10^{15}$ & $(1.60\pm 0.09)\times 10^{46}$ \\
    & $i^\prime$ & >$23.5$ & & $2.9\times10^{15}$ & $<2.6\times 10^{45}$ \\
    \\
    \multicolumn{6}{c}{X-ray:}\\
    \SRG/\eROSITA\ & 0.3--2~keV ($\Gamma=1.4$, unabsorbed) & $8.2^{+3.7}_{-2.7}\times 10^{-14}$~erg~cm$^{-2}$~s$^{-1}$ & 8 & & \\
    & 2--4~keV ($\Gamma=1.4$, unabsorbed) & $6.2^{+8.9}_{-3.9} \times 10^{-14}$~erg~cm$^{-2}$~s$^{-1}$ & & & \\
    \hline
    \multicolumn{3}{c}{Deduced quantities} & & & \\
    \hline
    \multicolumn{2}{l}{$\ar$ (140~MHz--1.4~GHz, observed)} & $\approx 0.5$ & & & \\
   \multicolumn{2}{l}{$\ar$ (1.6--5~GHz, observed)} & $\approx 1.0$ & & & \\
    \multicolumn{2}{l}{$\ar$ (5--32~GHz, observed)} & $\approx 0.7$ & & & \\
    \\ 
    \multicolumn{2}{l}{$\aosx$ (2500~\AA--2~keV)} & $1.12^{+0.57}_{-0.13}$  & & & \\
    \multicolumn{2}{l}{$\aox$ (2500~\AA--10~keV)} & $0.96^{+0.11}_{-0.07}$ & & & \\
    \multicolumn{2}{l}{$\nu L_\nu$ (2500~\AA)} & $ (1.9\pm0.3) \times 10^{46}$~erg~s$^{-1}$ & & & \\
    \multicolumn{2}{l}{$L$ (< 2~keV)} & $\gtrsim (1.2\pm 0.2) \times 10^{47}$~erg~s$^{-1}$ & & & \\
    \\
    \multicolumn{2}{l}{$\nu L_\nu$ (2~keV)} & $9.5_{-7.5}^{+32.7} \times 10^{45}$~erg~s$^{-1}$ & & & \\
    \multicolumn{2}{l}{$\nu L_\nu$ (10~keV)} & $2.5_{-1.3}^{+2.2} \times 10^{46}$~erg~s$^{-1}$ & & & \\
    \multicolumn{2}{l}{$L$ (2--10~keV)} & $2.6_{-1.0}^{+1.7} \times 10^{46}$~erg~s$^{-1}$ & & & \\
    \multicolumn{2}{l}{$L$ (2--30~keV)} & $6.5_{-2.5}^{+4.3} \times 10^{46}$~erg~s$^{-1}$ & & & \\
    \\
    \multicolumn{2}{l}{$L$ (bolometric)} & $\gtrsim 2\times 10^{47}$~erg~s$^{-1}$& & & \\
    \hline
\end{tabular}

\textbf{Notes.} Luminosities are given in the rest frame and are corrected for the Galactic extinction. The errors correspond to a 1 $\sigma$ confidence level, while the upper limits correspond to 5 $\sigma$.\\
\textbf{References.} (1) \cite{Shimwell_lofar}, (2) \cite{Condon_nvss}, (3) \cite{Helfand_first}, (4) \cite{Frey_VLBI}, (5) \cite{Wang_CO}, (6) \cite{Omont_fir}, (7) \cite{cutri13}, (8) this work, (9) \cite{ps1_survey_16}, (10) \cite{CFHQSIR}, (11) \cite{Hudelot_CFHTLS}. 

\end{table*}

\begin{figure*}
\centering
\includegraphics[width=0.9\textwidth]{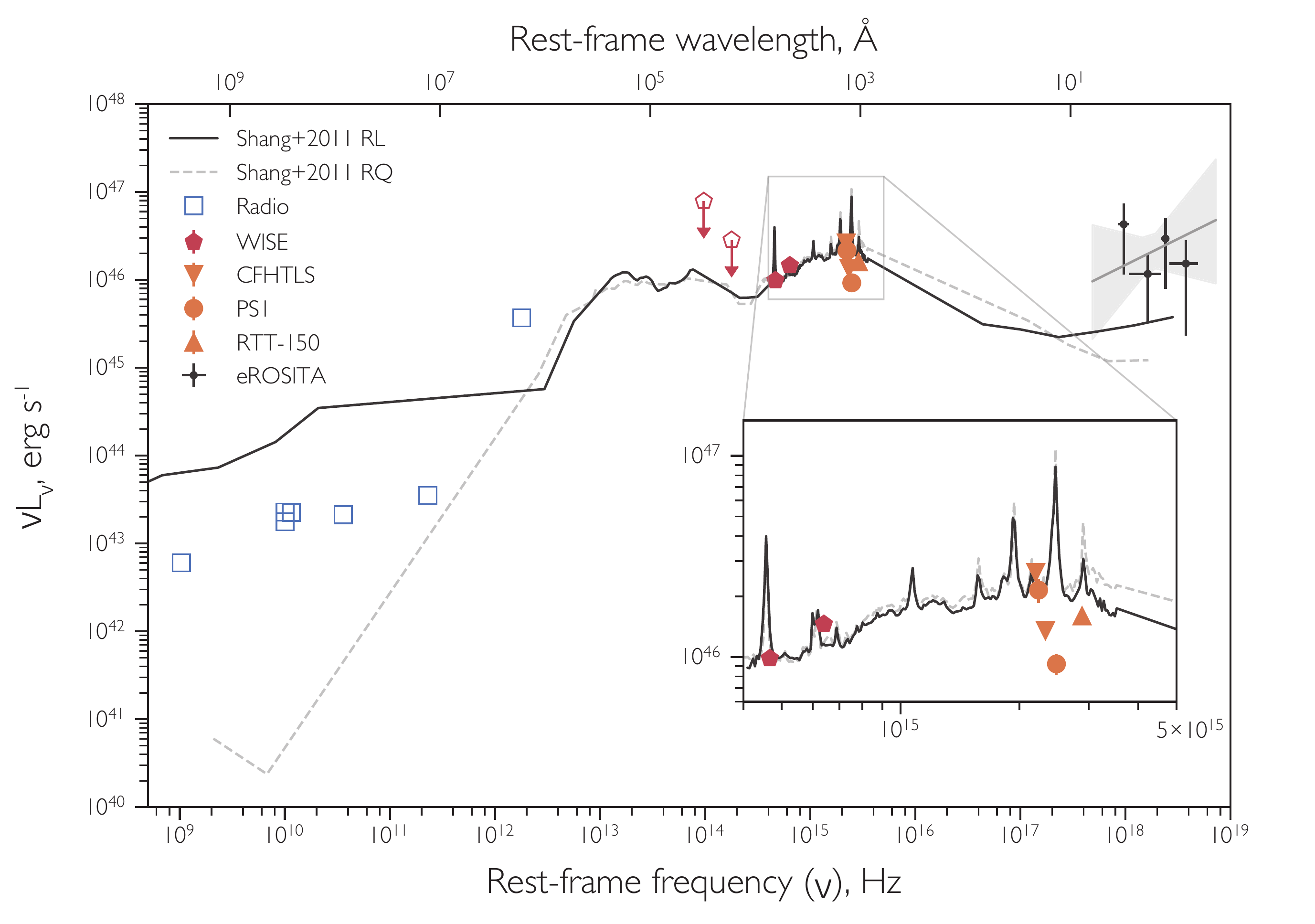}
\caption{Spectral energy distribution of \cfhqs\ (see Table~\ref{tab:sed} for details). 
The gray solid line with the associated $1\sigma$ uncertainty shows the X-ray part (rest-frame 2--30~keV) derived from the best fit of the \SRG/\eROSITA\ spectrum (black points) by an absorbed power-law model corrected for Galactic absorption. The optical data points are shown in orange (RTT150 by up-pointing triangles, CFHTLS by down-pointing triangles, and PS1 by circles). The infrared data points and upper limits (open symbols), resulting from \WISE\ forced photometry in the $W1$ and $W2$ bands on the PS1 position of \cfhqs\ and from the ALLWISE catalog in the $W3$ and $W4$ bands, are shown by the red pentagons. The blue squares show various radio measurements (the associated errors are smaller than the marker's size). Note that the inferred luminosity density at $1.8\times 10^{12}$~GHz (corresponding to the far-infrared band in the quasar's rest frame) is mostly associated with star formation activity in the host galaxy of \cfhqs\ rather than with SMBH growth in its nucleus \citep{Omont_fir}. For comparison, the median composite SEDs of radio-quiet and radio-loud quasars from \protect\cite{shang11}, normalized to the PS1 $y$-band flux density of \cfhqs, are shown by the dashed gray and the solid black lines, respectively. The inset shows a zoom-in on the optical part of the SED.}
\label{fig:sed}
\end{figure*}

Based on the multiwavelength (radio to X-ray) photometric measurements reported in Table~\ref{tab:sed}, we have constructed the spectral energy distribution (SED) of \cfhqs\ (Fig.~\ref{fig:sed}). In computing luminosities from the measured magnitudes in the optical band, we made a (small) correction for the Galactic extinction in the direction of the quasar, $A_V=0.05$ \citep{schlegel98}, assuming an extinction to reddening ratio $R_V=3.1$. 

\cfhqs\ is known to be radio-loud, with $R=109\pm 9$ (\citealt{Banados_radioloud_quasars}, see the definition of $R$ in \S\ref{s:intro}). A comparison of the SED of \cfhqs\ with the median composite SEDs of radio-quiet and radio-loud (non-blazar) quasars from \cite{shang11} confirms (see Fig.~\ref{fig:sed}) that it belongs to the radio-loud category, taking into account an order-of-magnitude scatter of the SEDs of individual RLQs in the radio and X-ray bands around the median SED (see \citealt{shang11}). Note that the sample used by \cite{shang11} is composed of quasars at $z\lesssim 1.5$, i.e. "nearby" objects compared to \cfhqs.

\subsection{X-ray--to--optical luminosity ratio}
\label{sec:aox}

What is revealed here for the first time is that apart from being radio-loud, \cfhqs\ is also X-ray bright in comparison to the majority of quasars. Indeed, its inferred X-ray luminosity is comparable to its luminosity at the expected peak (see e.g. \citealt{sazonov04}) of the SED near $\sim 1000$~\AA, with both being $\sim {\rm several}\times 10^{46}$~erg~s$^{-1}$ (see Fig.~\ref{fig:sed} and Table~\ref{tab:sed}). 

Quantitatively, X-ray brightness can be expressed in terms of the effective spectral slope between 2500~\AA\ and 2~keV \citep{tananbaum79},
\beq
\aosx\equiv -\frac{\log\Big(\Lsx/\Lo\Big)}{\log\Big(\nusx/\nuo\Big)}=-0.3838\,\log\left(\frac{\Lsx}{\Lo}\right),
\label{eq:aosx}
\eeq
or the corresponding slope between 2500~\AA\ and 10~keV \citep{Ighina_xray_blazars}, 
\beq
\aox\equiv -\frac{\log\Big(\Lx/\Lo\Big)}{\log\Big(\nux/\nuo\Big)}=-0.3026\,\log\left(\frac{\Lx}{\Lo}\right),
\label{eq:aox}
\eeq
where $\Lo$, $\Lsx$ and $\Lx$ are the luminosity densities (measured in units of erg~s$^{-1}$~Hz$^{-1}$) at rest-frame 2500~\AA, 2~keV and 10~keV, respectively. The $\aox$ is somewhat more convenient in application to $z\sim 6$ quasars than $\aosx$, since for such distant objects, 10~keV in the source's rest frame corresponds to $\sim 1.5$~keV in the observer's frame, an energy at which X-ray telescopes are most sensitive and the correction for line-of-sight absorption is relatively unimportant. 

At the redshift $z=6.183$ of \cfhqs, rest-frame 2500~\AA, 2~keV and 10~keV correspond to observed 1.8~$\mu$m, 0.3~keV and 1.4~keV, respectively. We estimated $\Lo$ using the median composite SED of radio-loud quasars from \cite{shang11}, normalized to the PS1 $y$-band flux density of \cfhqs\ (see Fig.~\ref{fig:sed}), while $\Lsx$ and $\Lx$ were determined from the best-fit spectral model describing the \eROSITA\ data in the 0.3--4~keV band (see \S\ref{s:xray}). Substituting the so derived values (see the bottom segment of Table~\ref{tab:sed}) into equation~(\ref{eq:aox}), we find $\aosx=1.11^{+0.25}_{-0.24}$ and $\aox=0.96_{-0.08}^{+0.10}$.

The X-ray--to--optical luminosity ratio for quasars has a long history of exploration, with most studies focused on RQQs. These usually find that $\aosx$ increases with increasing optical/UV luminosity (i.e. $\Lo$, e.g. \citealt{strateva05}). In particular, such a trend was reported by \cite{lusso10} based on a large X-ray selected sample of RQQs (see also \citealt{nanni17} for a confirmation of the same dependence at $z>5.5$), although the same authors inferred a nearly flat dependence $\aosx (\Lsx)$ (i.e. when the X-ray luminosity, rather than the optical one, was treated as an independent variable), with $\langle\aosx\rangle\approx 1.37$ and a dispersion of 0.18. Note that the RQQs studied by \cite{lusso10} have X-ray luminosities ranging between $\sim 10^{42}$ and $\sim 10^{45}$~erg~s$^{-1}$, while \cfhqs\ is much more X-ray luminous, having $\nu(2\,{\rm keV})\,\Lsx\sim 10^{46}$~erg~s$^{-1}$. 

In discussing the case of \cfhqs, it is more relevant to look at those studies that focus on RLQs. It is widely accepted that RLQs have relativistic jets, which are much weaker or absent in RQQs. The jets manifest themselves mainly in the radio but may also produce a significant amount of X-ray radiation in excess of the Comptonized emission from the hot corona of the accretion disk. This X-ray enhancement is expected to depend on the jet's orientation with respect to the observer and correlate with radio-loudness. \cite{miller11}, using a large sample of optically selected quasars, mostly at $z<4$, demonstrated that the $\aosx$ slope is indeed systematically flatter for RLQs compared to RQQs, although this effect was only noticeable at very high values of radio-loudness ($\log R\gtrsim 3.5$). For comparison, \cfhqs\ is only moderately radio-loud ($\log R\sim 2$) and the absolute majority of $z<4$ quasars of similar radio-loudness are less X-ray bright (with typical $\aosx$ values ranging between 1.6 and 1.2, see fig.~15 in \citealt{miller11}). In extension of that work, \cite{wu13} and \cite{zhu19} investigated the X-ray--to--optical luminosity ratio for $z\gtrsim 4$ quasars, concentrating on highly radio-loud objects ($\log R>2.5$). They confirmed that X-ray emission is enhanced in RLQs compared to RQQs and found evidence that the jet-linked X-ray excess is more significant in high-$z$ objects. This can at least partially be explained by the increased contribution of inverse Compton scattering of cosmic microwave background (CMB) photons in the jets, since the CMB energy density increases with redshift as $(1+z)^4$. Compared to $z>4$ RQQs of matched luminosity, the optical--to--X-ray slope typically flattens by $\Delta\aosx\sim -$(0.2--0.3) for $\log R\sim$ 2.5--3 (see fig.~5 in \citealt{zhu19}). Taking into account that \cfhqs\ is less radio-loud ($\log R\sim 2$) but is more distant ($z>6$) compared to the objects in the \cite{zhu19} sample ($z\sim $4--5), it seems plausible that its extreme X-ray brightness ($\aosx\sim 1.1$, or equivalently $\Delta\aosx\sim -0.5$) is associated with inverse Compton scattering of CMB photons in the jets.

Interestingly, the X-ray--to--optical luminosity ratio for \cfhqs, as e.g. expressed in terms of $\aox$, appears to be similar to typical ratios for $z>4$ blazars \citep{Ighina_xray_blazars}, although according to its radio properties \cfhqs\ is definitely not a blazar (see \S\ref{s:radio}). In particular, in the $\ar$--$R$ diagram (\citealt{Belladitta_blazar}, fig.~5), \cfhqs\ shares the same region with other presumably unbeamed high-$z$ RLQs, which is clearly separated from the region occupied by distant blazars (which all have $\ar\lesssim 0.5$ and typically have $R\sim 10^3$) including the most distant one, \pso, discussed in \S\ref{s:intro}. 


\subsection{Bolometric luminosity and SMBH mass}

We now wish to estimate the bolometric luminosity, $\Lbol$, of \cfhqs\ from its SED presented in Fig.~\ref{fig:sed}. The biggest uncertainty is posed by our ignorance of the intrinsic shape of the SED on the blue side of the Ly$\alpha$ transition (which approximately corresponds to the observed $z$ band) up to $\sim 2$~keV. Specifically, it is not clear how the optical--UV component of the SED, presumably associated with the emission from the accretion disk, connects to the X-ray component, which is likely the combined emission from the disk's corona and relativistic jets. 

We can first estimate the integrated luminosity of \cfhqs\ at energies below 2~keV. To this end, we can use the composite SED of RLQs from \cite{shang11} normalized to the PS1 $y$-band flux density of \cfhqs\ (see Fig.~\ref{fig:sed}). This yields $\Luv\sim 1.2\times 10^{47}$~erg~s$^{-1}$ with an uncertainty of $\sim 15\%$ (see Table~\ref{tab:sed}). However, although the \cite{shang11} template fits well the optical-UV part of the SED of \cfhqs, it probably underestimates the extreme-UV--soft-X-ray part of the SED. Therefore, the above estimate is likely a lower limit on $\Luv$. We also note that the discrepancy between the adopted template and the SED of \cfhqs\ in the radio band is not important here, since the radio emission contributes $\lesssim 1$\% of the bolometric luminosity. 

We can next estimate the luminosity at energies between 2 and 30~keV directly from the \SRG/\eROSITA\ data: $\Lbx\sim 6.5^{+4.3}_{-2.5}\times 10^{46}$~erg\,s$^{-1}$. However, as expected for AGNs (see e.g. \citealt{sazonov04, Malizia2014, Lohfink2017}), the X-ray spectrum of \cfhqs\ probably continues without a significant cutoff at least up to $\sim 100$~keV. A rough estimate of its total X-ray luminosity can thus be obtained by extrapolating our best-fit power-law model to 100~keV, which yields $\Lex= 1.5^{+3.5}_{-0.9}\times 10^{47}$~erg\,s$^{-1}$. The corresponding uncertainty is very large, since this estimate is currently based on a low-quality X-ray spectrum. If we fix the photon index at $\Gamma=1.8$ (a value typical for RQQs but not necessarily for RLQs), we find $\Lex= 1.0^{+0.4}_{-0.3 } \times 10^{47}$~erg\,s$^{-1}$.

We can finally estimate the bolometric luminosity of \cfhqs\ as $\Lbol=\Luv+\Lex\sim (2$--$3) \times 10^{47}$~erg\,s$^{-1}$. In reality, this may be a lower limit, since we may have underestimated the luminosity between $\sim 1000$\AA\ and 2~keV and above 100~keV. 

Assuming that the derived $\Lbol$ does not exceed the Eddington limit, we infer that the SMBH in \cfhqs\ must have a mass of least $\sim 1.5 \times 10^9$~$M_\odot$. 


\section{Discussion and conclusions}
\begin{figure}
\centering
\includegraphics[width=1\columnwidth]{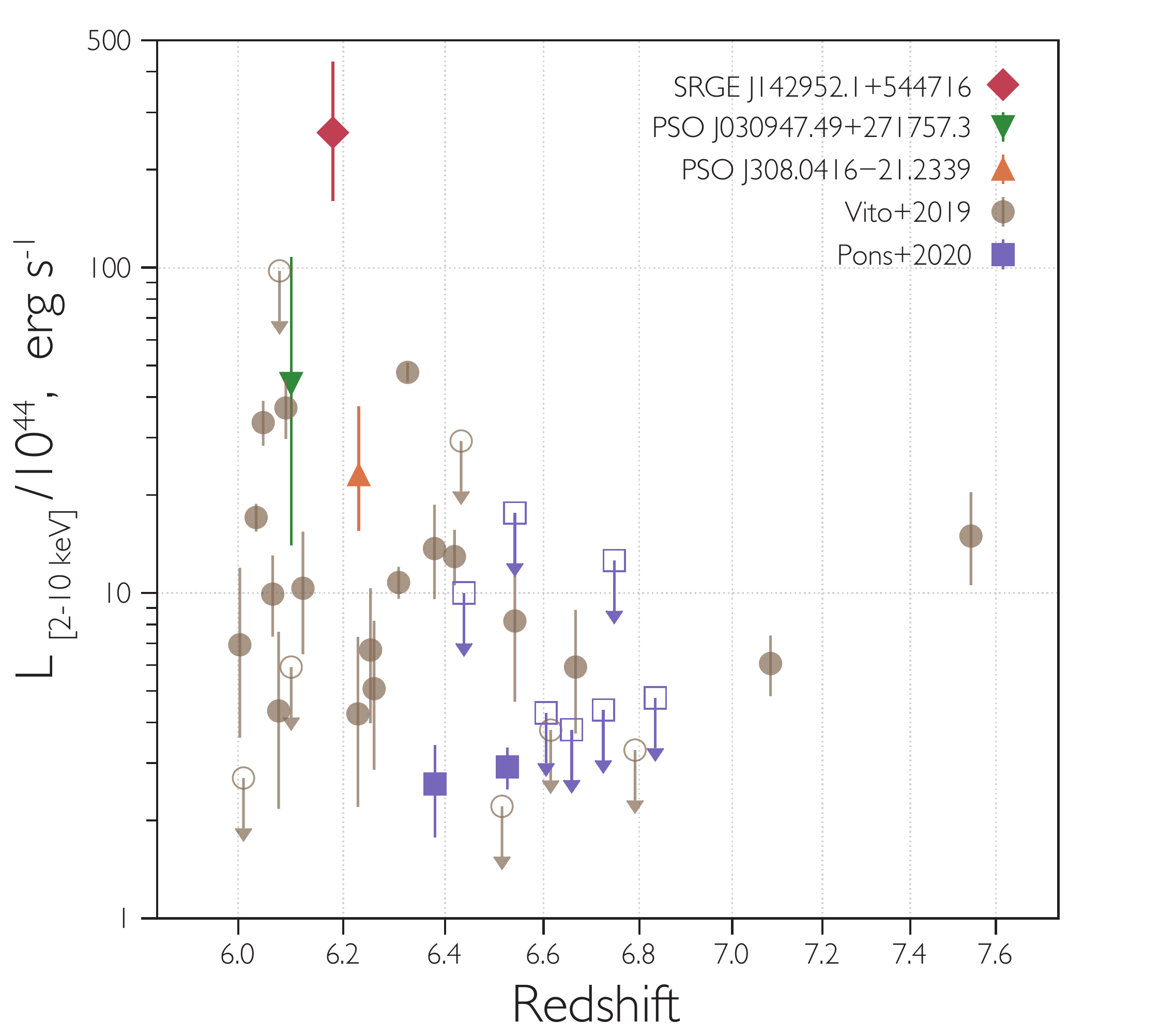}
\caption{X-ray luminosity in the rest-frame 2--10 keV band as a function of redshift for the $z>6$ quasars observed in X-rays so far. The brown circles are the 25 quasars from the compilation of  \protect\cite{Vito_xray_quasars}, the blue squares are the 9 quasars from the study by  \protect\cite{pons_quasars}, the orange triangle is the quasar PSO\,J308.0416$-$21.2339 \protect\citep{connor_quasars} and the green triangle is the blazar \pso\  \protect\citep{Belladitta_blazar}. The quasar \cfhqs\ analyzed in this work is shown by the red diamond. The open/filled symbols correspond to quasars not detected/detected in X-rays, respectively.}
\label{fig:z_Lx}
\end{figure}

With a luminosity of $2.6^{+1.7}_{-1.0}\times 10^{46}$~erg~s$^{-1}$ in the 2--10~keV energy band, \cfhqs\ proves to be the most X-ray luminous quasar at $z>6$ known so far. This is demonstrated in Fig.~\ref{fig:z_Lx}, where we show X-ray luminosity vs. redshift for all $z>6$ quasars with available X-ray measurements (detections or upper limits). This compilation is mainly based on the recent studies by \cite{Vito_xray_quasars} and \cite{pons_quasars}, and also includes the quasar PSO\,J308.0416$-$21.2339 \citep{connor_quasars} and the recently discovered blazar \pso\ \citep{Belladitta_blazar}. We see that even \pso, which already appeared extreme as regards its X-ray emission, is several times less X-ray luminous than \cfhqs, despite the fact that the latter is not a blazar according to its radio properties. 

In terms of bolometric luminosity, with $\Lbol\gtrsim 2\times 10^{47}$~erg~s$^{-1}$ \cfhqs\ is also among the most extreme quasars known at $z>5.7$ (see \citealt{matsuoka19} and in particular their fig.~3) but not the most luminous one. Hence, the most remarkable property of \cfhqs\ is its X-ray brightness (relative to the optical/UV emission). As we have discussed in section \ref{sec:aox}, it may be related to its radio-loudness, specifically to a plausible contribution of inverse Compton scattering of high-energy-density [due to the $(1+z)^4\approx 2.7\times 10^3$ cosmological factor] CMB photons off relativistic electrons in the jets. If so, \cfhqs\ might be just the tip of the iceberg of high-$z$ RLQs with enhanced X-ray emission, and \SRG/\eROSITA\ will probably find many more such objects (typically with somewhat lower luminosity) during its 4~year all-sky X-ray survey. 

On the other hand, we note that the inferred high X-ray brightness of \cfhqs\ may be partly associated with variability of the source. We recall that it was detected during its passage by \SRG/\eROSITA\ on Dec. 10--11, 2019, while all other existing multiwavelength observations are separated by at least several months from this date. In this connection, it will be interesting to check the X-ray activity of \cfhqs\ during its next passage by \SRG/\eROSITA, which is expected to occur in July 2020. 

\section*{Acknowledgements}
We are grateful to the reviewer for useful comments and suggestions.
This work is based on observations with eROSITA telescope onboard SRG observatory. The SRG observatory was built by Roskosmos in the interests of the Russian Academy of Sciences represented by its Space Research Institute (IKI) in the framework of the Russian Federal Space Program, with the participation of the Deutsches Zentrum für Luft- und Raumfahrt (DLR). The SRG/eROSITA X-ray telescope was built by a consortium of German Institutes led by MPE, and supported by DLR.  The SRG spacecraft was designed, built, launched and is operated by the Lavochkin Association and its subcontractors. The science data are downlinked via the Deep Space Network Antennae in Bear Lakes, Ussurijsk, and Baykonur, funded by Roskosmos. The eROSITA data used in this work were processed using the eSASS software system developed by the German eROSITA consortium and proprietary data reduction and analysis software developed by the Russian eROSITA Consortium.
The authors thank TÜBITAK, IKI, KFU, and AST for partial support in using RTT150 (the Russian-Turkish 1.5-m telescope in Antalya). AM acknowledge partial support of this work by the Russian Government Program of Competitive Growth of Kazan Federal University.  The work of IB and EI was partially funded by the subsidy 671-2020-0052 allocated to Kazan Federal University for the state assignment in the sphere of scientific activities. PM and SS acknowledge support from grant 19-12-00396 of the Russian Science Foundation.

\section*{Data availability}
X-ray data analysed in this article were used by permission of the Russian SRG/eROSITA consortium. The data will become publicly available as a part of the corresponding SRG/eROSITA data release along with the appropriate calibration information. Optical data used in the article will be shared on reasonable request to the corresponding author.

\bibliographystyle{mnras} 
\bibliography{distant_qso}

\end{document}